\documentclass[unnumsec,webpdf,contemporary,large]{oup-authoring-template}%

\graphicspath{{Fig/}}

\theoremstyle{thmstyleone}%
\theoremstyle{thmstyletwo}%
\theoremstyle{thmstylethree}%

\newcommand{\toolboxname}{{PROSO Toolbox}}

\begin{document}

\journaltitle{Journal Title Here}
\DOI{DOI HERE}
\copyrightyear{2023}
\pubyear{2023}
\access{Advance Access Publication Date: Day Month Year}
\appnotes{Applications Note}

\firstpage{1}


\title[\toolboxname]{\toolboxname: a unified protein-constrained genome-scale modelling framework for strain designing and optimization}

\author{Haoyang Yao\ORCID{0000-0002-7989-6529}}
\author[$\ast$]{Laurence Yang\ORCID{0000-0000-0000-0000}}

\authormark{Yao and Yang}

\address{\orgdiv{Department of Chemical Engineering}, \orgname{Queen's University}, \orgaddress{\street{19 Division Street, Kingston}, \postcode{K7L 2N9}, \state{ON}, \country{Canada}}}

\corresp[$\ast$]{Corresponding author. \href{email:laurence.yang@queensu.ca}{laurence.yang@queensu.ca}}

\received{Date}{0}{Year}
\revised{Date}{0}{Year}
\accepted{Date}{0}{Year}



\abstract{
The genome-scale metabolic model with protein constraint (PC-model) has been increasingly popular for microbial metabolic simulations.
We present \toolboxname, a unified and simple-to-use PC-model toolbox that takes any high-quality genome-scale metabolic reconstruction as the input.
The toolbox can construct a PC-model automatically, apply various algorithms for computational strain design and simulation, and help unveil metabolism from gene expression data through a state-of-the-art OVERLAY workflow. 
It also has detailed tutorials and documentation for maximum accessibility to researchers from diverse backgrounds.
\toolboxname, tutorials, and documentation are freely available online: \hyperlink{https://github.com/QCSB/PROSO-Toolbox}{https://github.com/QCSB/PROSO-Toolbox}.
}


\maketitle

\section{Introduction} \label{sec:intro}

For over two decades, genome-scale modelling (GEM) has been studied extensively as the mathematical twin of living microorganisms and cells, and it has drastically deepened our understanding of microbial decision-making. 
A mid-term goal of the GEM community is to achieve GEM-integrated synthetic biology through the "Design-Build-Test-Learn" (DBTL) cycle \cite{gray2018synthetic}. 
Genome-scale metabolic model (M-model), the most basic GEM framework, is a constraint-based mathematical network,
where each 
metabolite's mass balance
constitutes a linear constraint while metabolic reactions constitute a variable. 
COnstraint-Based Reconstruction and Analysis (COBRA) \cite{heirendt2019creation} has therefore been rapidly developing based on the M-model framework, thanks to its computational simplicity. 
Specifically, this simple constraint-based configuration is valid due to the generally 
fast time constants
of small-molecule-metabolites in the M-model. 
Yet the same does not apply to macromolecules including mRNA, proteins, and enzymes, all of which have 
slower concentration dynamics, and thus have not typically reached steady-state at the same time-scales as small-molecule metabolites.

A specialized GEM of metabolism and macromolecular expression (ME-model) has to be developed to encompass these missing aspects,
at the cost of fewer algorithms compatible and several orders of magnitude higher computational complexity \cite{lloyd2018cobrame}. 
This is especially problematic when applying GEM 
to interpret experiments where macromolecular mechanisms
are critical for the phenomena studied.

Protein-constrained metabolic model (PC-model) is a compromise between the M-model and ME-model, offering static and less detailed macromolecular insight without adding heavy computational burdens, and thus is gaining popularity in the community. 
Many GEM studies have adopted various PC-model formulations, and it is proven more effective than M-model \cite{zeng2020bridging,moreno2022enzyme}. 
However, due to a lack of a unified framework, PC-model studies have limited 
reusability
for new projects, which causes a general shortage of available algorithms for PC-model. 
In this study, we present a unified GEM framework with PROtein constraints for Strain designing and Optimization (\toolboxname). 
This toolbox features functions from our previous study, specifically automated protein constraint implementations and the state-of-the-art context-specific modelling algorithm OVERLAY \cite{yao2023novel}.
We have also adapted other algorithms from the COBRA community to fulfill the needs of a computational strain design, making it capable of completing the synthetic biology DBTL cycle. 

\section{Features} \label{sec:features}

\begin{figure*}[!t]%
\centering
\includegraphics[]{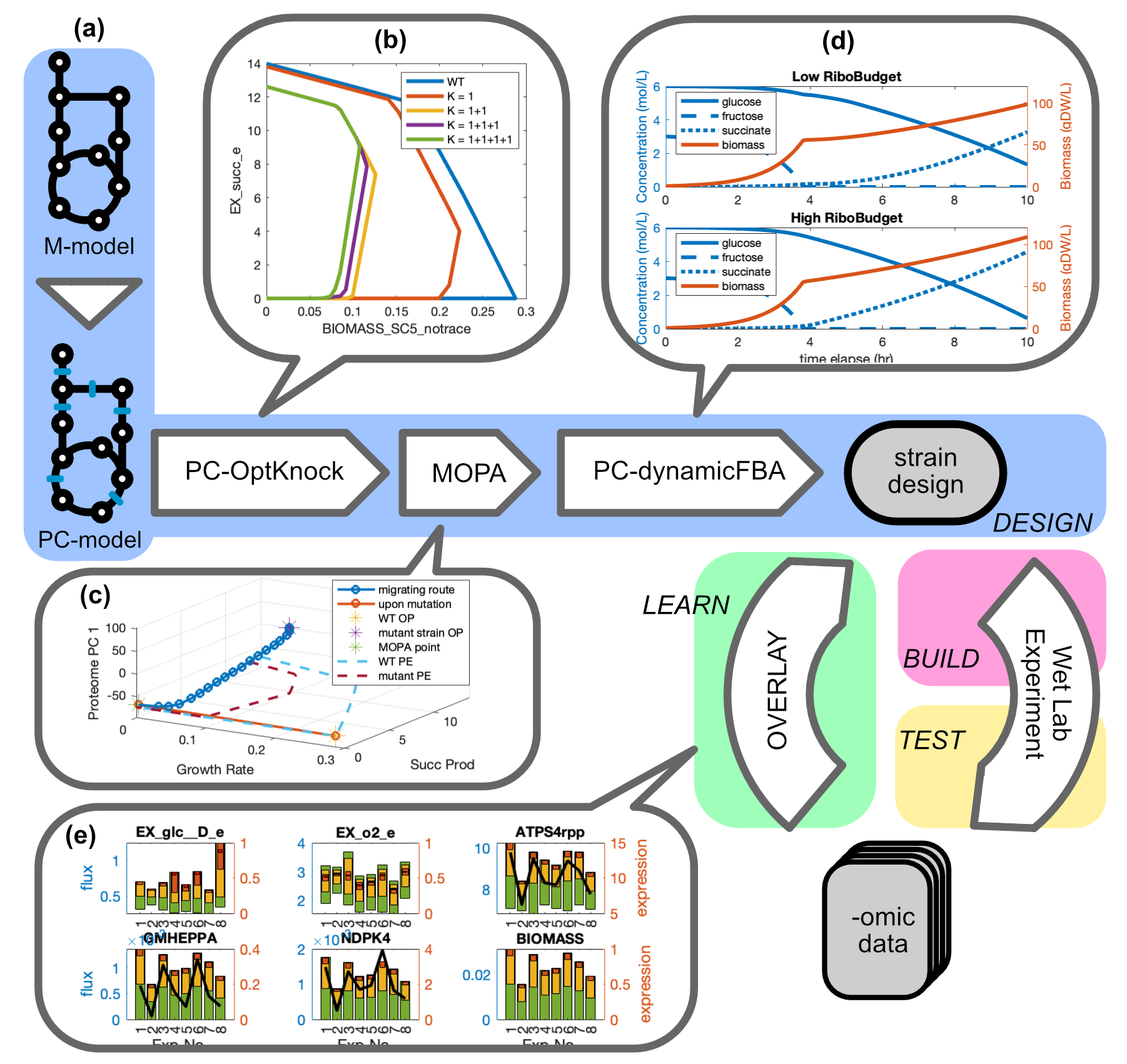}
\caption{
This is a conceptual schematic of the workflow and outputs offered by the \toolboxname, excluding the wet lab experiment. 
(\textbf{a}) M-model is used to construct a PC-model, of which PC-OptKnock, MOPA, and PC-dynamicFBA are applied for an \textit{in-silico} strain design. The design can be built and tested in the wet lab with -omic measurements. OVERLAY is then used to learn the underlying metabolism and update the design. This completes the DBTL cycle. 
(\textbf{b}) PC-OptKnock is used to design mutant strains with growth-coupled succinate productions. Production envelopes (succinate production versus biomass) for the wild-type and mutant strains are shown. 
(\textbf{c}) The algorithm of MOPA simulates the immediate effect (MOPA point), short-term effect (migrating route), and long-term effect (mutant strain OP) of mutations based on several assumptions. The relationship between these different states is plotted on the growth rate, succinate production rate, and proteome distance (proteome PC 1). The production envelopes (PE) for the wild-type and mutant strains are plotted in dotted lines. 
(\textbf{d}) Based on the results from previous steps, the consumption and accumulation of biomass and chemicals over multiple cell states can be simulated using PC-dynamicFBA. PC-dynamicFBA introduces ribosomal PC-FBA, which allows only limited proteome re-allocation over a certain time period according to a ribosomal budget. 
(\textbf{e}) OVERLAY is a multi-step pipeline that takes omic data as input and predicts cellular metabolism. The results from OVERLAY can help to refine the strain design for the next iteration of experiments.
}\label{fig1}
\end{figure*}

The \toolboxname \space consists of many smaller algorithms, such as PC-OptKnock, Minimization Of Proteomic Adjustments (MOPA), PC-dynamicFBA, and OVERLAY (Fig.~\ref{fig1}). 
The workflow takes an existing M-model as the input to formulate a PC-model, which is a metabolic model with protein constraints. 
This formulation of the PC-model will be a universal input for following strain design algorithms and OVERLAY, the previous of which in turn helps to propose and simulate strain designs. 
The \textit{in-silico} design can be built and tested in wet lab experiments, where multi-omic data such as transcriptomic, proteomic, and metabolomic measurements are taken. 
The pipeline of OVERLAY uses the PC-model and omic data as inputs to generate deep insights into cell metabolism, which in turn can help to replace modelling assumptions with knowledge as well as refine the strain design \cite{yao2023novel}. 
Some functions in \toolboxname \space require Gurobi Optimizer version 9.3 or above (offering free academic license) to solve non-convex bilinear optimizations \cite{gurobi}. 

\subsection{Protein constraints implementation} \label{subsec:pcmodel}

The \toolboxname \space features an automated protein constraint implementation as well as an optional and semi-automated curation step, most of which has been explained in our past publication \cite{yao2023novel}. 
Protein abundance is first added to the M-model as a variable. 
The toolbox then uses the gene-protein-reaction information in the metabolic reconstruction to add protein complex variables and couple them with respective metabolic reactions. 
The user may also choose to curate the protein complex subunit stoichiometry manually. 
The enzymatic rate constant can be tuned using surface area-based estimation, which results in a PC-model. 

\subsection{PC-OptKnock} \label{subsec:pcok}

The algorithm of PC-OptKnock is an adaptation of the original OptKnock onto the PC-model platform. 
In short, OptKnock is a 
bilevel optimization problem for designing
a microbial strain which, 
through repressing a small number of metabolic reactions, achieves a non-zero target chemical production rate when the microbe is growing optimally, also known as growth-coupling \cite{burgard2003optknock}. 
PC-OptKnock can suggest growth-coupling strategies by 
repressing
not reactions but proteins, which is more practical for wet lab scientists to implement and examine. 
Setting a larger maximum knockout number ($K$) is likely to achieve a higher chemical production rate at a significantly higher computational cost. 
Thus, we recommend the more computationally efficient method of conducting an iterative local search (i.e., $K=1+1+1$) than one larger global search (i.e., $K=3$) \cite{lun2009large}. 
The output from PC-OptKnock is comparable with OptORF \cite{kim2010optorf}, another growth-coupling strain designer that also encompasses multi-function enzymes and isozymes. 
Due to PC-OptKnock being based on the PC-model framework, it has higher prediction accuracy and more sensible output at a higher computational cost than the M-model-based OptORF. 

\subsection{MOPA} \label{subsec:MOPA}

The function of MOPA, inspired by the minimization of metabolic adjustment \cite{segre2002analysis},
finds the most likely cell state 
right after the mutation operation based on the proteomic proximity. 
Due to the speed of proteomic adjustment being orders of magnitude slower than the rate of metabolic or metabolomic adjustment, we believe MOPA is formulated on a better assumption that is made possible by our PC-model framework. 
The toolbox also features a subsequent function to compute the optimal proteomic adjustment path toward the new optimal state (Fig.~\ref{fig1}c, migrating route), of which the adjustment rate can be tuned. 

\subsection{PC-dynamicFBA} \label{subsec:pcdfba}

We design PC-dynamicFBA to integrate the productivity of the designated strain over time while tracking its proteomic profile. 
At each time step, the simulated cell needs to abide by the substrate availability and have a protein state proximal to the protein state of the previous time step. 
The maximum allowed rate of protein re-allocation in PC-dynamicFBA, or ribosome budget, determines the rate at which the cell adapts to a new condition, such as the depletion of a preferred substrate (Fig.~\ref{fig1}d). 
This also makes PC-dynamicFBA an enhanced framework for modelling diauxic growth compared to the original dynamic FBA \cite{mahadevan2002dynamic}. 

\subsection{OVERLAY: context-specific modelling} \label{subsec:overlay}

The pipeline of OVERLAY has been featured in our previous publication 
\cite{yao2023novel}. 
It is a streamlined workflow of two-step quadratic optimization (QP), debottlenecking, and protein-constrained flux variability analysis (PC-FVA) algorithms in order to unveil the metabolic insight behind gene expression data. 
Specifically, we used the first QP to fine-tune system-level enzymatic kinetics, and the second QP to implement data-specific protein constraints to PC-model. 
The subsequent two steps then exploit the plausible range of metabolic operations for the cell subject to the data-specific protein constraints and return to the user graphically. 
Fig.~\ref{fig1}e is a demonstration of the PC-FVA output, showing the predicted flux ranges of each metabolic reaction (bars) and its respective expression levels (lines). 
The nature of OVERLAY makes it a valuable tool for scientists to learn from the previous iteration of strain design and experiment. 

\subsection{Other utilities} \label{subsec:utility}

Apart from the functions listed above, the \toolboxname \space has other utility functions for easy debugging, understanding, and further development. 
For example, the function \textit{tryCloseRxn.m} helps the user quickly set up exchange fluxes of an M-model by solving a mixed integer optimization; 
the function \textit{gb2faa.m} parses an NCBI GenBank file into a proteomic FASTA file that is compatible with the toolbox; 
the function \textit{minimalGenome.m} is a stand-alone algorithm to find the minimal number of active proteins that is growth sustaining; 
\textit{biocycUtility} allows the user to download flat files from the BioCyc database \cite{karp2019biocyc} and map items to the PC-model in a semi-automatic fashion. 
We believe these utility functions will help to serve users and make their access and contribution easier. 

\section{Comparison with other toolboxes} \label{sec:benchmark}

The method of constraining metabolic networks with proteomic or enzymatic allocation has been exploited by many other studies. 
The most well-known workflows are MOMENT \cite{adadi2012prediction} and GECKO \cite{sanchez2017improving}, both featuring extensive details on enzyme kinetics. 
GECKO 2.0 \cite{domenzain2022reconstruction} is the most detailed enzymatic framework to date and is claimed to be comparable to the ME-model. 
Consequently, 
they require more data inputs and potentially complex procedures (possibly requiring more domain expertise or computational resources) to estimate additional model parameters.

On the other hand, simplified versions of PC-model (or equivalently, ec-model) have been implemented in other studies, i.e., sMOMENT \cite{bekiaris2020automatic}, ECM \cite{noor2016protein}, and ECMpy \cite{mao2022ecmpy}. 
These packages are easy to use, usually without adding enzymes as variables, yet their condensed form makes them less versatile for further improvements and integrating omic data. 

In comparison, \toolboxname \space is automated and accessible without sacrificing its versatility in incorporating details and omic data. 
We allow users to automatically construct a draft PC-model with no additional input on enzyme or protein information, which has proven to enhance the prediction compared to the base M-model. 
The user can also choose to supply data on protein sequence, protein complex stoichiometry, effective enzymatic turnover rate, and proteomic weight fraction to further optimize the performance of PC-model. 
We have adapted many algorithms for the M-model and FBA into our toolbox, demonstrating the versatility of our framework for further development. 
More importantly, the framework is compatible with OVERLAY to
algorithmically
interpret transcriptomic or proteomic data through the PC-model,
which provides a distinctive functionality in our toolbox.
We believe the \toolboxname \space has closed the DBTL cycle, and thus it is especially suitable for the purpose of computational strain design and optimization for synthetic biologists. 

\section{Conclusion} \label{sec:conclusion}

In this study, we introduced the \toolboxname, a PC-model framework that unifies automatic protein constraint implementation, strain designing and simulating functions, and OVERLAY as the algorithmic expression data interpreter. 
As for the PC-model framework, it is uniquely capable of computational strain designs for wet lab implementation, as well as taking the transcriptomic data measured from the experiment as input for further optimization. 
We believe the \toolboxname \space will strongly bridge the GEM community with wet lab experiments and help to achieve GEM-integrated synthetic biology. 

\section{Competing interests}

No competing interest is declared.

\section{Author contributions statement}

H.Y. and L.Y. conceptualized the project. 
H.Y. implemented and tested the algorithms.
H.Y. wrote the documentation. 
H.Y. drafted the manuscript. 
H.Y. and L.Y. revised the manuscript.
L.Y. directed the research. 

\section{Acknowledgments}

The first author thanks Ziying Wang for her assistance in testing and proofreading. 
This work was funded by the Government of Canada through Genome Canada and Ontario Genomics (OGI-207), the Government of Ontario through an Ontario Research Fund (ORF), and Queen’s University.

\bibliographystyle{plain}
\bibliography{reference}

\end{document}